\documentclass[fleqn,12pt,twoside]{article}
\usepackage{espcrc1}
\usepackage{times}

\usepackage{graphicx}
\usepackage[figuresright]{rotating}

\newcommand{\AmS}{{\protect\the\textfont2
  A\kern-.1667em\lower.5ex\hbox{M}\kern-.125emS}}
\title{ Chiral and 
$U(1)_A$ restorations high in the hadron spectrum,
 semiclassical approximation and large $N_c$.}
\author{{ L. Ya. Glozman}\vskip5mm Institute for Theoretical
Physics, University of Graz, Universit\"atsplatz 5, A-8010
Graz, Austria\footnote{e-mail: leonid.glozman@uni-graz.at}}
\begin{document}
\maketitle

\begin{abstract} 
In quantum systems with large $n$ (radial quantum number) 
or large angular momentum the semiclassical (WKB)
approximation is valid. A physical content of the
semiclassical approximation is that the quantum
fluctuations effects 
are suppressed and vanish asymptotically.
The chiral as well as $U(1)_A$ breakings in QCD  result
from quantum fluctuations. Hence these breakings must be
suppressed high in the spectrum and the spectrum of high-lying
hadrons must exhibit $U(2)_L \times U(2)_R$ symmetry of the classical QCD
Lagrangian. This argument can be made stronger for mesons in the large
$N_c$ limit. In this limit all mesons are stable
against strong decays and the spectrum is infinite. Hence,
one can excite  mesons of arbitrary large size with arbitrary
large action, in which case the semiclassical limit is manifest.
Actually we do not need the exact $N_c=\infty$ limit. For any large
action there always exist such $N_c$ that the isolated mesons with
such an action do exist and can be described semiclassically. From
the empirical fact that we observe multiplets of chiral and
$U(1)_A$ groups high in the hadron spectrum it follows that
$N_c=3$ is large enough for this purpose. 

\end{abstract}

\bigskip
\bigskip

\section{Introduction}
If one neglects the tiny masses of $u$ and $d$ quarks, which
are much smaller than $\Lambda_{QCD}$, then the QCD Lagrangian
exhibits the

\begin{equation}
U(2)_L\times U(2)_R = SU(2)_L\times SU(2)_R\times U(1)_V\times U(1)_A
\label{sym}
\end{equation}

\noindent
symmetry. This is because the quark-gluon interaction
Lagrangian in the chiral limit does not mix the left- and
right-handed components of quarks and hence the total
QCD Lagrangian for the two-flavor QCD can be split into
the left-handed and right-handed parts which do not communicate
with each other. We know that the $U(1)_A$ symmetry of the
classical QCD Lagrangian is absent at the quantum level
because of the $U(1)_A$ anomaly, which is an effect
of quantum fluctuations \cite{ANOMALY}. We also know
that the chiral $SU(2)_L\times SU(2)_R$ symmetry is spontaneously
(dynamically)
broken in the QCD vacuum \cite{NJL}. That this is so is directly evidenced
by the nonzero value of the quark condensate, 
$\langle \bar q q \rangle = \langle \bar q_L q_R + \bar q_R q_L \rangle
\simeq -(240 \pm 10 MeV )^3$, which represents an order
parameter for spontaneous chiral symmetry breaking. This
quark condensate  shows that
in the QCD vacuum the left-handed quarks are correlated
with the right-handed antiquarks (and vice verca) and
hence the QCD vacuum breaks the chiral symmetry.
This spontaneous (dynamical) breaking of chiral symmetry
is  a pure quantum effect based upon quantum fluctuations. To see the latter
we remind the reader that the chiral symmetry
breaking can be formulated via the Schwinger-Dyson (gap)
equation. It is not yet clear at all which specific gluonic
interactions are the most important ones as a kernel of the
Schwinger-Dyson equation (e.g. instantons
\cite{SS}\footnote{
The instanton itself is an Euclidean semiclassical gluon field configuration.
But chiral and $U(1)_A$ symmetry breakings by instantons is
a quark field quantum fluctuations process. This is because
the 't Hooft effective interaction is obtained only upon
integrating of the $U(1)_A$ anomaly.}
, or 
gluonic exchanges \cite{A}, or perhaps  other gluonic
interactions, or a combination of different interactions).
But in any case the quantum fluctuations  of the quark
and gluon fields are
very strong in the low-lying hadrons and induce both 
chiral and $U(1)_A$ breakings. As a consequence we do not
observe any chiral or $U(1)_A$ multiplets low in the hadron
spectrum.

That the spontaneous breaking of chiral symmetry is an effect
of quantum fluctuations of the quark field can be seen most
generally from the definition of the quark condensate,
which is a closed quark loop:

\begin{equation}
\langle \bar \psi \psi \rangle =
- Tr \lim_{x \rightarrow 0_+}
\langle 0 | T \left \{ \psi(0) \bar \psi(x) \right \} | 0 \rangle.
\label{cond}
\end{equation}

\noindent
This closed quark loop explicitly contains a factor $\hbar$. The
chiral symmetry breaking, which is necessaraly a nonperturbative
effect, is actually a (nonlocal) coupling of a quark line with the
closed quark loop, which is a tadpole graph. Hence it always
contains an extra factor $\hbar$ as compared to the tree-level
quark line.

\begin{figure}
\begin{center}
\includegraphics*[width=5cm,angle=-90]{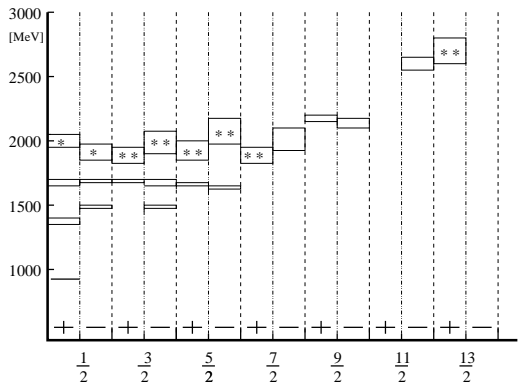}
\end{center}
\caption{Excitation spectrum of the nucleon. The real part
of the pole position is shown. Boxes represent experimental
uncertainties. Those resonances which are not yet established
are marked by two or one stars according to the PDG classification.
 }
\end{figure}

The upper part of both baryon \cite{G1,CG1} and
meson \cite{G2,G22} spectra almost systematically 
exhibits multiplets of the chiral and $U(1)_A$ groups
(for a pedagogical overview see \cite{G3}),
though a careful experimental exploration 
of high-lying spectra must
be done for a final conclusion. This phenomenon
is referred to as effective chiral symmetry restoration
or chiral symmetry restoration of the second kind. 
This is illustrated
in Fig. 1 and Fig. 2, where the excitation spectrum of
the nucleon from the PDG compilation \cite{PDG} as well as the
excitation spectrum of $\pi$ and $f_0$ (with the $\bar n n = \frac{\bar u u +
\bar d d}{\sqrt 2}$ content) mesons \cite{G2} are shown.
\footnote{Those scalar mesons which are not $\bar n n$ states,
and hence irrelevant for the present analysis, have been
removed from the consideration. These are well established
 $f_0(1500)$ and $f_0(1710)$,
 which are believed to be mostly glueball and $\bar s s$ states,
 respectively, as well as $f_0(2102)$ which is seen in $\bar p p$
 and considered by the authors of the partial wave analysis as
 a glueball due to its decay modes \cite{A2,BUGG}.}

Starting from the 1.7 GeV region the nucleon spectrum shows
obvious signs of parity doubling. There are couple of examples
where chiral partners of highly excited states have not yet been
seen. Their experimental discovery would be an important
task. All possible irreducible representations of the 
parity-chiral group necessarily contain parity doublets \cite{CG1}.
Similarly, in the chirally restored regime $\pi$ and $\bar n n$ $f_0$
states must be systematically degenerate.

\begin{figure}
\begin{center}
\includegraphics*[width=7cm,angle=-90]{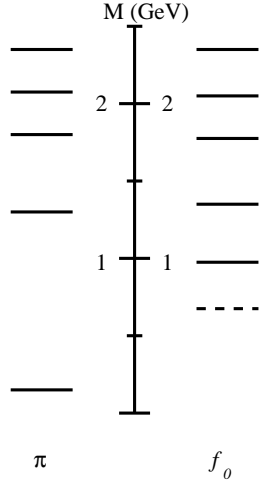}
\end{center}
\caption{Pion and $n \bar n$ $f_0$ spectra.}
\end{figure}

A quantitative measure of chiral symmetry breaking 
contribution to the hadron mass at the leading (linear) order
is  the {\it chiral asymmetry}

\begin{equation}
\chi = \frac{|M_1 - M_2|}{(M_1+M_2)},
\label{chir}
\end{equation}

\noindent
where $M_1$ and $M_2$ are masses of particles within the
same multiplet. This parameter  has the interpretation of the part
 of the hadron mass  due to chiral
symmetry breaking. From the low-lying states the chiral
asymmetry is typically 0.2 - 0.6, as can be seen e.g.
from comparison of masses of the states $1/2^+,N(939)$ and $1/2^-,N(1535)$,
$\pi$ and $\sigma$,   $\rho(770)$
and $a_1(1260)$, or   $\rho(770)$ and $h_1(1170)$. If 
the chiral asymmetry is large as above, then it makes no
sense to assign a given hadron to the chiral multiplet
since its wave function is a strong mixture of different
representations and we have to expect also large
{\it nonlinear} symmetry breaking effects.
Indeed, there is even no one-to-one mapping of the positive
and negative parity hadrons low in the spectrum. It indicates
that here the chiral symmetry breaking effects are very strong
and as a consequence the chiral symmetry is realized nonlinearly.

\begin{table}
\begin{center}
\caption{Chiral multiplets of $\pi$ and $\bar n n ~~~f_0$  mesons
\cite{G2}.
Comments: (i) $\pi(1300)$ and $f_0(1370)$ are well established states
and can be found in the Meson Summary Table of Review of Particle 
Physics \cite{PDG}. (ii) $\pi(1812 \pm 14)$ is abbreviated as
$\pi(1800)$ in the Meson Summary Table of Review of Particle 
Physics \cite{PDG}; $f_0(1770 \pm 12)$ is seen in $\bar p p$
as $\bar n n$ state \cite{A1,BUGG} and also recently as $15 \sigma$
peak in the $\pi \pi$ channel in
$J/\Psi$ decays at $\sim 1790$ MeV \cite{BES}; The analysis
of ref. \cite{CZ} confirms a $\bar n n$ nature of this state.
(iii) These states are clearly seen in a few different channels
in $\bar p p$ \cite{A2,BUGG},
though in order to appear in the Meson Summary Table of Review of Particle 
Physics they must be reconfirmed by independent experiment.
 }
 \begin{tabular}{|lllll|} \hline
Chiral multiplet &  Representation & $\chi$ & Spectral Overlap & Comment\\ \hline
$\pi(1300 \pm 100) - f_0(1370^{+130}_{-170})$ & (1/2,1/2) &
$0.03^{+0.09}_{-0.03}$  & $0.1^{+0.3}_{-0.1}$  & (i) \\

$\pi(1812 \pm 14) - f_0(1770 \pm 12)$ & (1/2,1/2) &
$0.012 \pm 0.007$  & $0.09 \pm 0.06$  & (ii) \\

$\pi(2070 \pm 35) - f_0(2040 \pm 38)$ & (1/2,1/2) &
$0.007^{+0.017}_{-0.007}$  & $0.11^{+0.28}_{-0.11}$  & (iii) \\

$\pi(2360 \pm 25) - f_0(2337 \pm 14)$ & (1/2,1/2) &
$0.005^{+0.009}_{-0.005}$  & $0.08^{+0.13}_{-0.08}$  & (iii) \\

\hline
\end{tabular}
\end{center}
\end{table}
\begin{table}
\begin{center}
\caption{Chiral multiplets of high-lying $J=2$ mesons with the
u,d valence quark content \cite{G22}. 
Comments: 
(i) These states are clearly seen in a few different channels
in $\bar p p$ \cite{A2,A3,BUGG},
though in order to appear in the Meson Summary Table of Review of Particle 
Physics they must be reconfirmed by independent experiment.
 }
\begin{tabular}{|lllll|} \hline
Chiral multiplet &  Representation & $\chi$ & Spectral Overlap & Comment\\ \hline
$\omega_2(1975 \pm 20) - f_2(1934 \pm 20)$ & (0,0) &
$0.01 \pm 0.01$  & $0.16 \pm 0.15$  & (i) \\

$\omega_2(2195 \pm 30) - f_2(2240 \pm 15)$ & (0,0) &
$0.01 \pm 0.01$  & $0.17 \pm 0.17$  & (i) \\

$\pi_2(2005 \pm 15) - f_2(2001 \pm 10)$ & (1/2,1/2) &
$0.001^{+0.006}_{-0.001}$  & $ 0.02^{+0.09}_{-0.02}$  & (i) \\

$\pi_2(2245 \pm 60) - f_2(2293 \pm 13)$ & (1/2,1/2) &
$0.01^{+0.02}_{-0.01}$  & $ 0.18^{+0.27}_{-0.18}$  & (i) \\

$a_2(2030 \pm 20) - \eta_2(2030 \pm ?)$ & (1/2,1/2) &
$0.0 \pm ?$  & $0.0 \pm ?$  & (i) \\

$a_2(2255 \pm 20) - \eta_2(2267 \pm 14)$ & (1/2,1/2) &
$0.003^{+0.008}_{-0.003}$  & $ 0.05^{+0.015}_{-0.05}$  & (i) \\

$a_2(1950^{+30}_{-70} ) - \rho_2(1940 \pm 40)$ & $(0,1) \oplus (1,0)$  &
$0.003^{+0.018}_{-0.003}$  & $ 0.04^{+0.27}_{-0.04}$  & (i) \\

$a_2(2175  \pm 40 ) - \rho_2(2225 \pm 35)$ & $(0,1)\oplus (1,0)$  &
$0.011^{+0.017}_{-0.011}$  & $ 0.20^{+0.29}_{-0.20}$  & (i) \\

\hline
\end{tabular}
\end{center}
\end{table}

 However, recent experimental data show that at meson 
masses about 1.8 - 2.3  GeV the chiral multiplets persist
systematically (with a few missing states which should
be discovered) and
the chiral asymmetry is  within 0.01 \cite{G2,G22}.

A useful parameter that characterizes a "goodness" of symmetry
in the spectrum is the {\it spectral overlap} which is defined as
a ratio of the splitting witin the multiplet to the distance
between centers of gravity of two subsequent multiplets. Clearly,
the symmetry is "good" and can be easily recognized
 if this parameter is much smaller than 1.

Both chiral asymmetries and spectral overlaps as well as
assignements of mesons with $J=0$ and $J=2$, where the data
sets are complete enough, are given in Tables 1 and 2.\footnote
{Note that the Table 2 also evidences a restoration
of $U(1)_A$ symmetry \cite{G2,G22,G3}. To see the latter one
has to compare masses of states from the distinct
(1/2,1/2) multiplets with the same isospin and opposite
parity: $\pi_2 - a_2$, $f_2 - \eta_2$. The chiral asymmetries
and spectral overlaps in these cases are similar to those ones
in Table 2.}

In
the nucleon spectrum the chiral asymmetry is smaller than 0.02
for approximate parity doublets in the region $ M \sim 1.7$ GeV.
Similar doublets are observed in the delta spectrum starting
from the $ M \sim 1.9$ GeV. This means that the parity doubling
in both cases is seen at approximately the same excitation energy
with respect to the corresponding ground state. This indicates that
not an absolute value of the energy is important in order to
approach the chiral symmetry restoration regime but rather
radial quantum number $n$ or angular momentum $J$. Chiral multiplets
in the nucleon spectrum are listed in Table 3. 

\begin{table}
\begin{center}
\caption{Chiral multiplets of excited nucleons
\cite{CG1}.
Comments: (i) All these states are well established and
can be found in the Baryon Summary Table of Review of Particle 
Physics. 200 MeV is taken as an interval between the consequent 
multiplets in order to evaluate the spectral overlap. (ii) There
are two possibilities to assign the chiral representation:
$(1/2,0) \oplus (0,1/2)$ or $(1/2,1) \oplus (1,1/2)$ because
there is a possible chiral pair in the $\Delta$ spectrum
with the same spin with similar mass.
 }
\begin{tabular}{|llllll|} \hline
Spin & Chiral multiplet &  Representation & $\chi$ & Spectral Overlap & Comment\\ \hline
1/2& $N_+(1710 \pm 30) - N_-(1650^{+30}_{-10})$ & $(1/2,0) \oplus (0,1/2)$ &
$0.02 \pm 0.02$  & $0.3 \pm 0.3$  & (i) \\

3/2& $N_+(1720^{+30}_{-70}) - N_-(1700^{+50}_{-50})$ & $(1/2,0) \oplus (0,1/2)$ &
$0.01^{+0.03}_{-0.01}$   &  $0.1^{+0.4}_{-0.1}$ & (i) \\

5/2&$N_+(1680^{+10}_{-5}) - N_-(1675^{+10}_{-5})$ & $(1/2,0) \oplus (0,1/2)$ &
$0.002^{+0.006}_{-0.002}$   &  $0.025^{+0.1}_{-0.025}$ & (i) \\

9/2&$N_+(2220^{+90}_{-40}) - N_-(2250^{+60}_{-80})$ &
 see comment (ii) &
$0.01^{+0.03}_{-0.01}$   &  $0.15^{+0.75}_{-0.15}$ & (i),(ii) \\

\hline
\end{tabular}
\end{center}
\end{table}
 
 In all these cases the hadrons can be believed to be  members
of  chiral multiplets with a tiny admixture of other
representations. It also means that practically the whole
mass of the hadron {\it is not related} to the chiral
symmetry breaking in the vacuum.

\section{Chiral symmetry restoration in excited hadrons by definition}

By definition 
 effective symmetry restoration means the following. In QCD 
hadrons with  quantum numbers $\alpha$ are created
when one applies the local interpolating field (current) $J_\alpha(x)$
with such quantum numbers  on the vacuum 
$|0\rangle$. Then all 
hadrons that are created by the given interpolator
appear as intermediate states in the two-point correlator

\begin{equation}
\Pi_{J_\alpha}(q) = i \int d^4x~ e^{-iqx}
\langle 0 | T \left \{ J_\alpha(x) J_\alpha(0)^\dagger\right \} | 0 \rangle,
\label{corr}
\end{equation}

\noindent
where all possible Lorentz and Dirac indices ( specific for
a given interpolating field) have been omitted.
Consider two local interpolating fields  $J_1(x)$ and 
$J_2(x)$ which are connected by a chiral transformation,

\begin{equation}
J_1(x) = U J_2(x) U^\dagger,
\end{equation}

\noindent
where 

\begin{equation}
U \in SU(2)_L \times SU(2)_R
\end{equation}

\noindent
(or by the $U(1)_A$ transformation). Then if the vacuum was
invariant under the chiral group, 

\begin{equation}
U|0\rangle = |0\rangle,
\end{equation}

\noindent
it would follow from (\ref{corr}) that the spectra created by the
operators  $J_1(x)$ and  $J_2(x)$ would be identical.

 We know that in QCD  one finds

\begin{equation}
U|0\rangle \neq |0\rangle.
\end{equation}

\noindent
 As a consequence the spectra of the two operators
must be in general different 
 and we do not observe any chiral
or $U(1)_A$ multiplets in the low-lying hadron spectra.
 However,
it happens that the noninvariance of the vacuum becomes
unimportant (irrelevant) high in the spectrum. Then the masses
of the corresponding opposite parity hadrons, which 
are the members of the given parity-chiral multiplet,
become close  at large $s$ 
(and identical asymptotically high),

\begin{equation}
M_1 - M_2 \rightarrow 0.
\end{equation}

We stress that this effective 
chiral symmetry
restoration does not mean that chiral symmetry breaking in
the vacuum disappears, but  that the role of the quark
condensates that break chiral symmetry in the vacuum becomes progressively
less important high in the spectrum. One could say that the valence
quarks in high-lying hadrons {\it decouple} from the QCD vacuum.\\

\section{Restoration of the classical symmetry in the
semiclassical regime}

In ref. \cite{CG1} a justification for this chiral symmetry
restoration  has been suggested. Namely,
at large space-like momenta
$Q^2 = -q^2 > 0$ the correlator
can be adequately represented by the operator product
expansion, where all nonperturbative effects reside in
different condensates \cite{SVZ}. The only effect that
spontaneous breaking of chiral symmetry can have on the
correlator is via the quark condensate of the vacuum,
$\langle \bar q  q \rangle$, and higher dimensional
condensates that are not invariant under chiral transformation $U$.
However, the contributions of all these condensates are suppressed
by the inverse powers of momenta $Q^2$.  This shows that
at large space-like momenta the correlation function
becomes chirally symmetric. 
The dispersion relation provides a connection between the
space-like and time-like domains  of the correlator. In particular,
the large $Q^2$ correlator is completely dominated by the
large $s$ spectral density $\rho(s)$, which is an observable.
Hence the large $s$ spectral density should be insensitive
to the chiral symmetry breaking in the vacuum and the spectra
of two operators $J_1(x)$ and  $J_2(x)$ should approach each other,

\begin{equation}
\rho_1(s) - \rho_2(s) \rightarrow 0, ~~~~ s \rightarrow \infty.
\end{equation}

\noindent
This is in contrast to the low $s$ spectra
 which are very different
because of the chiral symmetry breaking in the vacuum.

Unfortunately OPE at large $Q^2$ does not allow us
to make quantitative statements concerning the
functional behaviour that determines approaching
the chiral-invariant regime at large $s$. This is because
the OPE is only an asymptotic expansion and hence
cannot be continued to the time-like region.

While the argument above on the asymptotic symmetry
properties of spectral functions is rather robust 
(it is based actually only on
the asymptotic freedom of QCD at large space-like momenta
and on the analyticity of the two-point correlator), 
{\it a-priori} it
is not clear whether it can be applied to the bound state 
systems, which the hadrons are. Indeed, it can happen that
the asymptotic symmetry restoration applies only to that
part of the spectrum, which is above the resonance region
 (i.e. where the current creates jets but not isolated hadrons). So
the question arises whether it is possible to prove (or at
least justify) the symmetry restoration in highly excited
{\it isolated} hadrons. We show below that both chiral
and $U(1)_A$ restorations in highly excited isolated hadrons  can
be anticipated as a direct consequence of the semiclassical
regime in the highly excited hadrons, indeed.

At large $n$ (radial quantum number) or at large angular
momentum $J$ we know that in quantum systems the {\it semiclassical}
approximation (WKB) {\it must} work. Physically this approximation
applies in these cases because the de Broglie wavelength of
particles in the system is small in comparison with the
scale that characterizes the given problem. In such a system
as a hadron the scale is given by the hadron size while the
wavelength of valence quarks is given by their momenta. Once
we go high in the spectrum the size of hadrons increases as well as
 the typical momentum of valence quarks.
This is why a highly excited hadron  can be described semiclassically
in terms of the underlying quark  degrees of freedom.

\begin{figure}
\begin{center}
\includegraphics*[width=10cm]{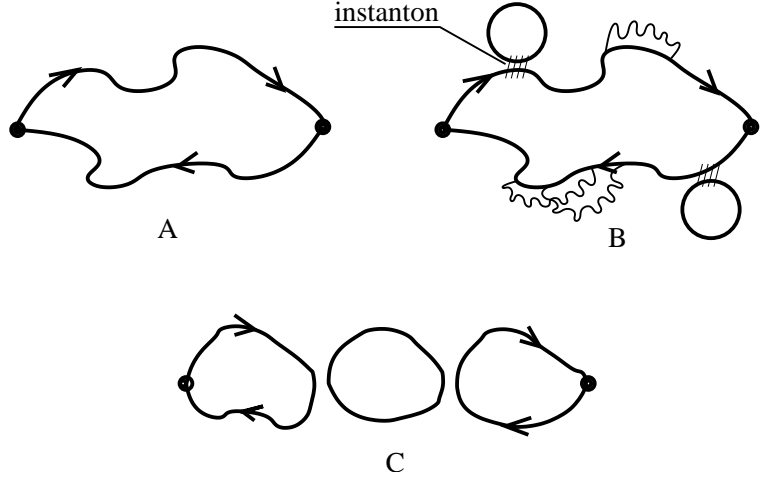}
\end{center}
\caption{Tree level (A)  and typical quantum fluctuations
contributions (B and C) to the two-point function.}
\end{figure}

The physical content of the semiclassical approximation is
most transparently given by the path integral. The contribution
of the given path to the path integral is regulated by the
action $S(\phi(x))$ along the  path $\phi(x)$ (the
fields $\bar \psi,\psi,A$ are collectively denoted as $\phi$) through
the factor

\begin{equation}
\sim e^{i\frac{S(\phi(x))}{\hbar}}.
\end{equation}

\noindent
The semiclassical approximation  applies when the
action in the system $S \gg \hbar$.
In this case the whole amplitude (path integral) is dominated by
the classical path $\phi_{cl}(x)$ (stationary point) and those paths that 
are infinitesimally
close to the classical path. All other paths that differ from
the classical one by an appreciable amount  do
not contribute. These latter paths would represent the quantum fluctuation
effects. In other words, in the semiclassical case the quantum
fluctuations effects are strongly suppressed and vanish asymptotically.
Mathematically it follows from the well known statement that
when $\hbar \rightarrow 0$  the functional integral can
be calculated in the stationary phase approximation (in Euclidean
space it is a steepest descent or saddle point approximation).
The stationary point is a solution of the classical equations
of motion in the presence of the source $J(x)$,

\begin{equation}
\frac{\delta S}{\delta \phi(x)}[\phi_{cl}(J)] = - J(x).
\label{cl}
\end{equation}

\noindent
Then the generating functional can be expanded
in powers of $\hbar$ as

\begin{equation}
W(J) = W_0(J) + \hbar W_1(J) + ...,
\label{f}
\end{equation}

\noindent
where $W_0(J) = S(\phi_{cl}) + J\phi_{cl}$ and $W_1(J)$
represents contributions of the lowest order quantum fluctuations around the
classical solution (determinant of the classical
solution).\footnote{The expansion (\ref{f}) should not be mixed
with the quenched QCD. While there are no vacuum fermion loops
in quenched QCD like those in Fig. 3c, there are still quantum
fluctuations of the valence quark lines  which induce
chiral symmetry breaking, see Fig.3b. The first term in the expansion
(\ref{f}) does not contain any quantum fluctuations of the
quark fields.}

The classical path, which is generated by $W_0$,
 is a tree-level contribution
to the path integral, see diagram a) on Fig. 3,
 and keeps chiral symmetries of the classical
Lagrangian. Its contribution is of the order
$(\hbar/S)^0$. The quantum fluctuations contribute at the
orders $(\hbar/S)^1$ (the one loop order, generated by $W_1$),
 $(\hbar/S)^2$ (the two loops order),
etc, see graphs b) and c) on Fig. 3.

The $U(1)_A$ as well as the spontaneous $SU(2)_L \times SU(2)_R$
breakings  result from quantum fluctuations of the quark fields
 and start from the
one-loop order.
However, in a hadron with
large enough $n$ or  $J$, where action is large,  the quantum fluctuations
contributions must be relatively
suppressed and vanish asymptotically.
Then it follows 
that in such systems both
the chiral and $U(1)_A$ symmetries must be restored.
This is precisely what we see
phenomenologically. In the nucleon spectrum the doubling
appears either at large $n$ excitations of baryons with
a given small spin or in resonances of large spin. Similar
features persist in the delta spectrum. In the meson spectrum
the doubling is obvious for large $n$ excitations of small
spin mesons  \cite{G2} and there are signs of doubling of large spin
mesons (the data are, however, sparse). It would be certainly
interesting and important to observe systematically multiplets
of parity-chiral and parity-$U(1)_A$ groups  (or, sometimes, when
the chiral and $U(1)_A$ transformations connect {\it different}
hadrons, the multiplets of the
$U(2)_L \times U(2)_R$ group \cite{G2} ). The high-lying hadron spectra
must be systematically explored.

While the argument above is solid, theoretically it is
not clear {\it a-priori} whether isolated hadrons still
exist at excitation energies where a semiclassical regime
is achieved. Hence it is conceptually important to demonstrate that
in QCD hadrons still exist, while the dynamics inside such
 hadrons already is semiclassical. We do not know how to prove it for
$N_c =3$. However, the large $N_c$ limit of QCD \cite{H},
while  keeping all basic properties of QCD like
asymptotic freedom and chiral symmetry, allows for a significant
simplification. It is a good approximation, e.g., for the low-lying
baryons \cite{M}.

In this limit it is known that all mesons
represent  narrow states, i.e. they are stable against
strong decays. At the same time the spectrum of mesons is
infinite (see, e.g., \cite{W}). The latter is necessary to match
the two-point function in the perturbation theory regime
(which contains logarithm)
at large space-like momenta with the discrete spectral sum in the
dispersion integral. Then one can always excite a meson of any
 arbitrary large energy, which is of any arbitrary large size.
 In such a meson the action $S \gg \hbar$. Hence a description
 of this meson necessarily must  be semiclassical. Then the equation
 of motion in such a meson must be according to some yet
 unknown solution of the classical QCD Lagrangian
for a colorless meson. Hence
 it must exhibit chiral and $U(1)_A$ symmetries.  
  This proves that it is possible to have an isolated hadron
 which can be described semiclassically.

Actually we do not need the exact $N_c = \infty$ limit
for this statement. It can be formulated in the following
way. For any large $S \gg \hbar$ there always exist such
$N_c$ that the isolated meson with such an action does exist
and can be described semiclassically. From the  empirical
fact that we observe  multiplets of chiral
and $U(1)_A$ groups high in the hadron spectrum it follows
that $N_c=3$ is large enough for this purpose.                                                                    
 
The strength of the argument given above is that it is very
general. However we
cannot say anything concrete about 
how all this happens. For that one needs a detailed microscopical
understanding of both confinement and chiral symmetry breaking
 in QCD, and in particular a classical solution of the QCD Lagrangian
 for a highly-excited color-singlet hadron, around which an
 expansion (13) can be performed,
  which is a challenging task. But even though we can only
 assume 
how microscopically all this happens, it is a solid statement
 that in highly excited
hadrons symmetries of the classical QCD Lagrangian should be observed.
The only basis for this statement is that in such hadrons a semiclassical
description is correct.
\footnote{That the quantum 
fluctuations effects vanish in  quantum bound state systems
at large $n$ or $J$ 
is well known e.g. from the Lamb shift. The Lamb shift is a result
of the radiative corrections (which represent
effects of quantum fluctuations of electron and electromagnetic
fields)
 and vanishes  as $1/n^3$,
and also very fast with increasing $J$. As a consequence high in the
hydrogen spectrum  the symmetry of the classical Coulomb
potential gets restored. The author is grateful to D.O. Riska
for suggesting this nice analogy.
The other well-known example 
is the 't Hooft model \cite{HM} (QCD in 1+1 dimensions).
In this model in the regime $N_c \rightarrow \infty$, 
$m_q \rightarrow 0$, $m_q \gg g \sim 1/\sqrt N_c$, the spectrum
of the high-lying states is known exactly, $M_n^2 \sim n$. The
chiral symmetry of the Lagrangian is broken (with no contradiction
with the Coleman theorem since in this specific regime everything
is determined by $N_c = \infty$ ,
 for any large but finite $N_c$ the chiral symmetry is not 
 broken in agreement with the Coleman
theorem ), which is due to gluon dressing of valence quarks and
 which is reflected by the fact that the positive
and negative parity states are not degenerate and alternate
in the spectrum. However, the mass difference between the
 neighbouring positive and negative parity states is
$M_+ - M_- \sim 1/\sqrt n$ and  vanishes high in the spectrum
since the effect of quantum fluctuations dies out
high in the spectrum. 
The latter can be explicitly seen from the fact that the
amplitude of the higher quark Fock component in the meson wave function
dies out very fast with increasing n \cite{Li}.
The author is grateful to T. Cohen, T. DeGrand,
S. Peris and A. Zhitnitsky for discussions on this point.
}

\section{Conclusions}

Using very general arguments we have shown that both
$SU(2)_L \times SU(2)_R$ and $U(1)_A$ symmetries of
the classical QCD Lagrangian should be approximately
restored in  highly excited hadrons and manifest
asymptotically high. The reason is that in  highly excited
hadrons physics must necessarily be semiclassical.
  Unfortunately
 solutions of the classical QCD equations of motion for
 the white hadrons  are yet unknown. Hence we can  make only
 an assumption about the corresponding physical picture. If 
one constructs a highly excited hadron as a string (flux tube)
with quarks at the ends that have definite chirality \cite{G4},
which is a semiclassical picture, then one necessarily obtains
all hadrons in chiral and $U(1)_A$ multiplets. Such a picture
is rather natural and is well compatible with the Nambu
string picture \cite{N}. The ends of the string in the
Nambu model move with velocity of light. Then, (it is
an extention of the Nambu model) the quarks at the ends
of the string must have definite chirality. In this way one
is able to explain at the same time both Regge trajectories,
chiral multiplet structure of excited hadrons and absense of
the spin-orbit force in the $u,d$ sector.

\bigskip

The work was supported by the FWF project P16823-N08
of the Austrian Science Fund.

\end{document}